\documentclass[aps,pre,preprint,groupedaddress,showpacs,showkeys]{revtex4-1}
\usepackage{graphicx}
\usepackage{amssymb, amsmath}

\begin{document}

\title{Scaling properties of correlated random walks}

\author{Claus Metzner}
\email[]{claus.metzner@gmx.com}
\affiliation{Biophysics Group, University of Erlangen, Henkestr.91, D-91052 Erlangen, Germany}

\date{\today}

\begin{abstract}
Many stochastic time series can be modelled by discrete random walks in which a step of random sign but constant length $\delta x$ is performed after each time interval $\delta t$. In correlated discrete time random walks (CDTRWs), the probability $q$ for two successive steps having the same sign is unequal $1/2$. The resulting probability distribution $P(\Delta x,\Delta t)$ that a displacement $\Delta x$ is observed after a lagtime $\Delta t$ is known analytically for arbitrary persistence parameters $q$. In this short note we show how a CDTRW with parameters $\left[  \delta t, \delta x, q \right]$ can be mapped onto another CDTRW with rescaled parameters $\left[  \delta t/s, \delta x\cdot g(q,s), q^{\prime}(q,s) \right]$, for arbitrary scaling parameters $s$, so that both walks have the same displacement distributions $P(\Delta x,\Delta t)$ on long time scales. The nonlinear scaling functions $g(q,s)$ and $q^{\prime}(q,s)$ and derived explicitely. This scaling method can be used to model time series measured at discrete sample intervals $\delta t$ but actually corresponding to continuum processes with variations occuring on a much shorter time scale $\delta t/s$. 

\end{abstract}



\maketitle




\section{The dimensionless CDTRW}

Originally, a CDTRW does not have any physical scales $\delta x$ or $\delta t$ attached to its steps, but uses dimensionless integers $k$ and $n$ to count the discrete units of space and time. The trajectory of the fictuous random walker (the sequence of integer `positions' $X_n$) is determined by the initial position $X_0$ (usually unimportant) and the sequence of steps, or differences $\nu_n=X_n-X_{n-1}$. These steps $\nu_n$ only take two possible values $-1$ and $+1$ and can thus be called sign factors. Since the probabilty of each sign depends only on the directly preceeding one, a discrete correlated random walk is a Markov chain. It is therefore easy to show that for a given persistence probability $q$, the autocorrelation function of $\nu_n$ is given by 
\begin{equation}
C_{\nu\nu}(n\;|\;q) = \frac{ \left\langle \nu_m\;\nu_{m+n}\right\rangle_m } { \left\langle \nu^2_m\right\rangle_m } = (2q-1)^n.
\label{css}
\end{equation}

Since the resulting statistical properties of the CDTRW model have been derived analytically in Ref.\cite{hanneken98}, we just repeat the main results here. The displacement distribution for a DTCRW with persistence parameter $q$ is given by 
\begin{eqnarray}
&P&(k,n\;|\;q) = \nonumber\\ &=& \sum_{m=1}^{(n-|k|)/2}\! \binom{(n\!+\!k\!-\!2)/2}{m\!-\!1} \binom{(n\!-\!k\!-\!2)/2}{m\!-\!1} \nonumber\\ &\cdot& (1-q)^{2m-1}q^{n-1-2m}  \left( \frac{n(1\!-\!q)+2m(2q\!-\!1)}{2m}  \right),
\end{eqnarray}
where $n$ and $k$ must either both be even or both be odd. It has additionally been assumed that initially (at time step $t\!=\!0$) the probablity for the particle to go left or right are equal. Note that $P(k,n)=0$ for $|k|>n$. The corresponding mean squared displacement, defined as the second moment of $P(k,n\;|\;q)$, is given by
\begin{eqnarray}
&&\overline{\Delta k^2}(n\;|\;q) =\nonumber\\
&&\frac{nq}{1-q}\left\{ 1-\frac{(2q-1)\left[1-(2q-1)^n\right]}{2nq(1-q)} \right\}.
\end{eqnarray}

\section{The CDTRW with physical dimensions}

In order to relate the CDTRW model to an actual measured time series, it is neccessary to associate the integers $k$ and $n$ with physical quantities (having dimensions) by defining $x_k=k\;\delta x$ and $t_n=n\;\delta t$. Then, the `physical' displacement distribution is given by

\begin{equation}
p(\Delta x,\Delta t \;|\;q,\delta x,\delta t) = P(k\!=\!\Delta x/\delta x, n\!=\!\Delta t/\delta t \;|\;q) / (2\delta x).
\end{equation}

\begin{figure}[htb]
\includegraphics[width=10cm]{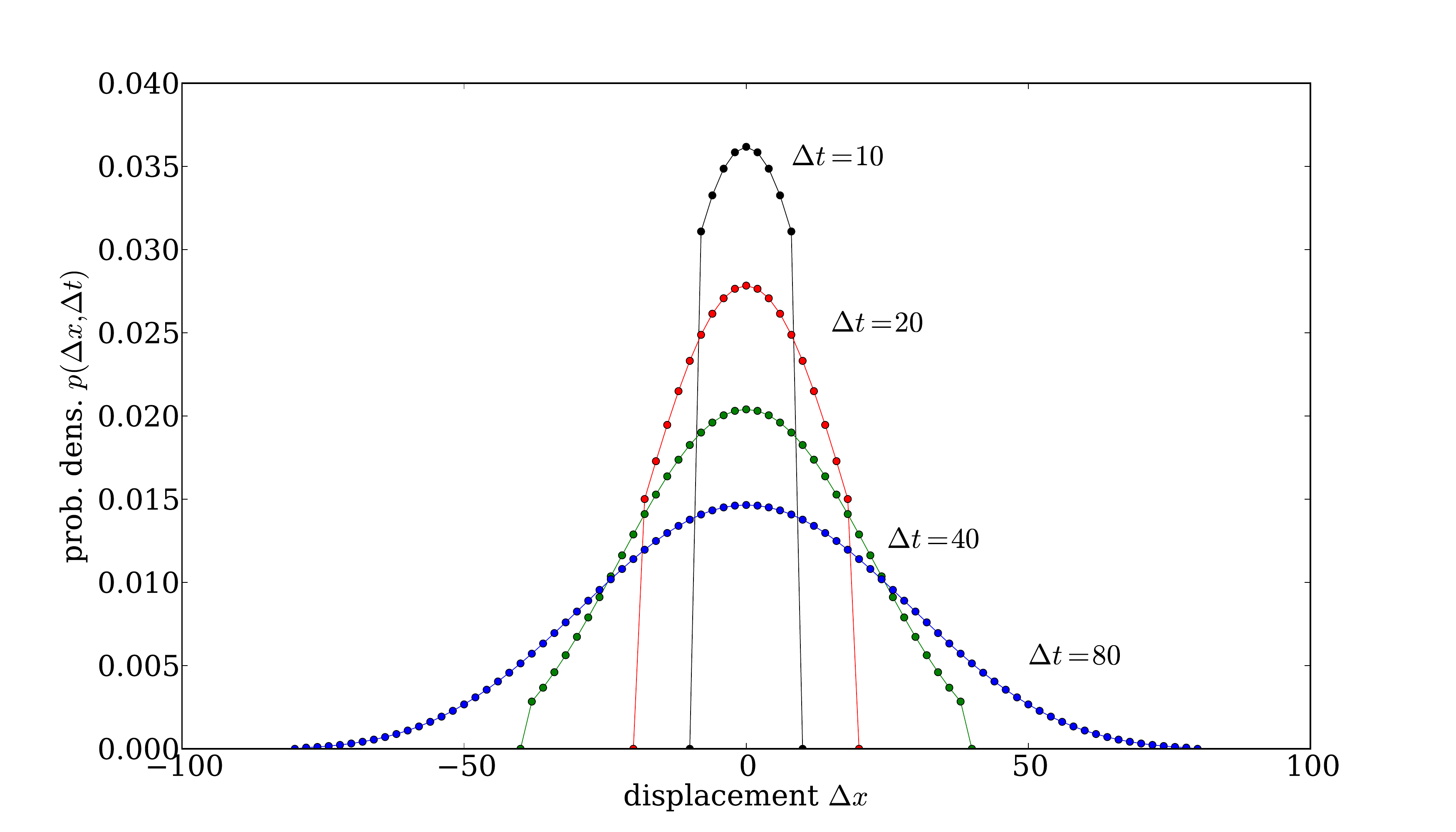}
\caption{\label{fig:swdQ09} {\em Displacement distributions for a CDTRW model with parameters $\delta x=1$, $\delta t=1$ and $q=0.9$. Note that the probability density is zero for $\Delta x > \Delta t$. The dots correpond to the discrete values of the model. The lines are interpolated.}}
\end{figure}

Displacement probabilities for $\Delta x$ and $\Delta t$ that are not integer multiples of $\delta x$ and $\delta t$ can be obtained by interpolation (compare Fig.\ref{fig:swdQ09}).

The physical mean squared is given by

\begin{equation}
\overline{\Delta x^2}(\Delta t\;|\;q,\delta x,\delta t) = \overline{\Delta k^2}(n\!=\!\Delta t/\delta t\;|\;q) \cdot \delta x^2.
\end{equation}

Again, function values for non-integer multiples of $\delta t$ can be interpolated.

\section{Scaling of the persistence probability}

\begin{figure}[htb]
\includegraphics[width=10cm]{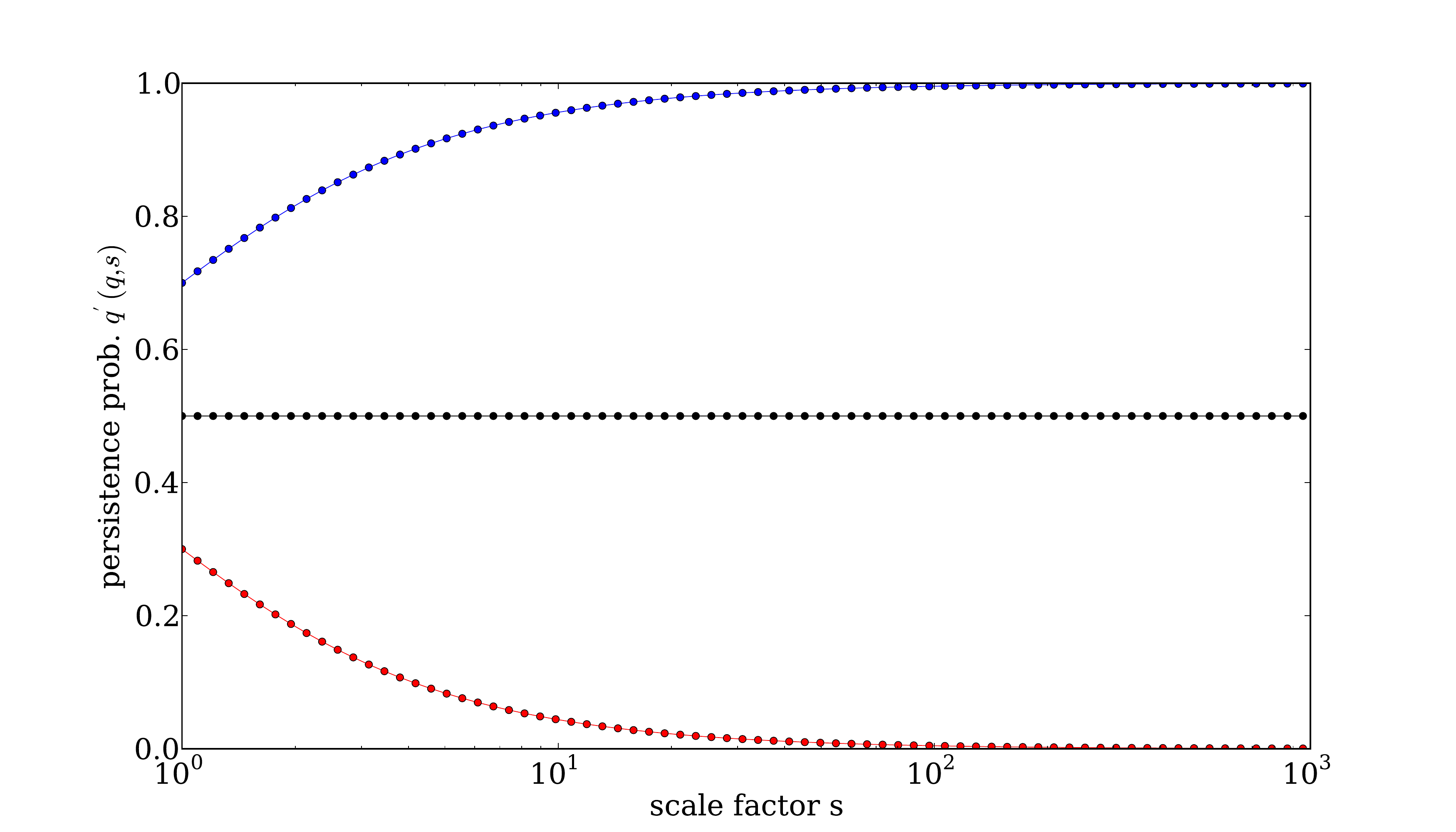}
\caption{\label{fig:flux} {\em Dependence of the rescaled persistence probability $q^{\prime}(q,s)$ on the scaling parameter $s$ for the original $q$-values $0.3$ (red), $0.5$ (black) and $0.7$ (blue). }}
\end{figure}

Consider a correlated random walk {\em A} with a persistence parameter $q>\frac{1}{2}$. We seek to construct another walk {\em B} that after $s\cdot n$ steps has the same correlations as {\em A} after $n$ steps. We therefore require 
\begin{equation}
C_{\nu\nu}(n\;|\;q)=C_{\nu\nu}(s\cdot n\;|\;q^{\prime}).
\end{equation}
Inserting Eq.(\ref{css}) and solving for the unknown persistence parameter $q^{\prime}$ of walk {\em B} yields $q^{\prime}= \frac{1}{2}\left( 1+\left| 2q-1  \right|^{1/s}  \right)$. In the case of persistence parameters $q<\frac{1}{2}$ we can write $(2q-1)^n$ as $(-1)^n |2q-1|^n$ and perform an analogous calculation. As a result, we obtain a scaling rule for the persistence parameter:  
\begin{equation}\boxed{
q^{\prime}=q^{\prime}(q,s) = \frac{1}{2}\left( 1\pm \left| 2q-1  \right|^{1/s}  \right)},
\end{equation}
where the sign $+$ ($-$) has to be used if $q>1/2$ ($q<1/2$). This transformation has no effect on uncorrelated walks with $q=0.5$. Persistent walks with $q>0.5$ become even more persistent. Anti-persistent walks with $q<0.5$ become even more anti-persistent (compare Fig.\ref{fig:flux}).

\section{Scaling of the step length}

While the above transformation of the persistence probability ensures that the orignal walk {\em A} and the rescaled walk {\em B} have equivalent autocorrelation functions, this does not automatically guarantee equivalence of the mean squared displacements. In order to achive the latter point, the step length has to be rescaled as well. For this purpose, note that for lagtimes $\Delta t$ much longer than the correlation time, the mean squared displacement behaves like that of a diffusive process and is given by $\overline{\Delta x^2}\rightarrow \delta x^2 \;q(1-q)\;\Delta t/\delta t$. To match the assymptotic values of the two walks, we therefore require
\begin{equation}
\overline{\Delta x^2}(s\cdot\Delta t\;|\;q^{\prime},\delta x^{\prime},\delta t)=\overline{\Delta x^2}(\Delta t\;|\;q,\delta x,\delta t).
\end{equation}
As a result, we obtain an additional scaling law for the step length:
\begin{equation}\boxed{
\delta x^{\prime} = \delta x\cdot g(q,s)\;\;\mbox{with}\;\; g(q,s) = \sqrt{\frac{q(1-q)}{q^{\prime}(1-q^{\prime})} \frac{1}{s}  }}.
\end{equation}

\section{Summary and Examples}

\begin{figure}[htb]
\includegraphics[width=10cm]{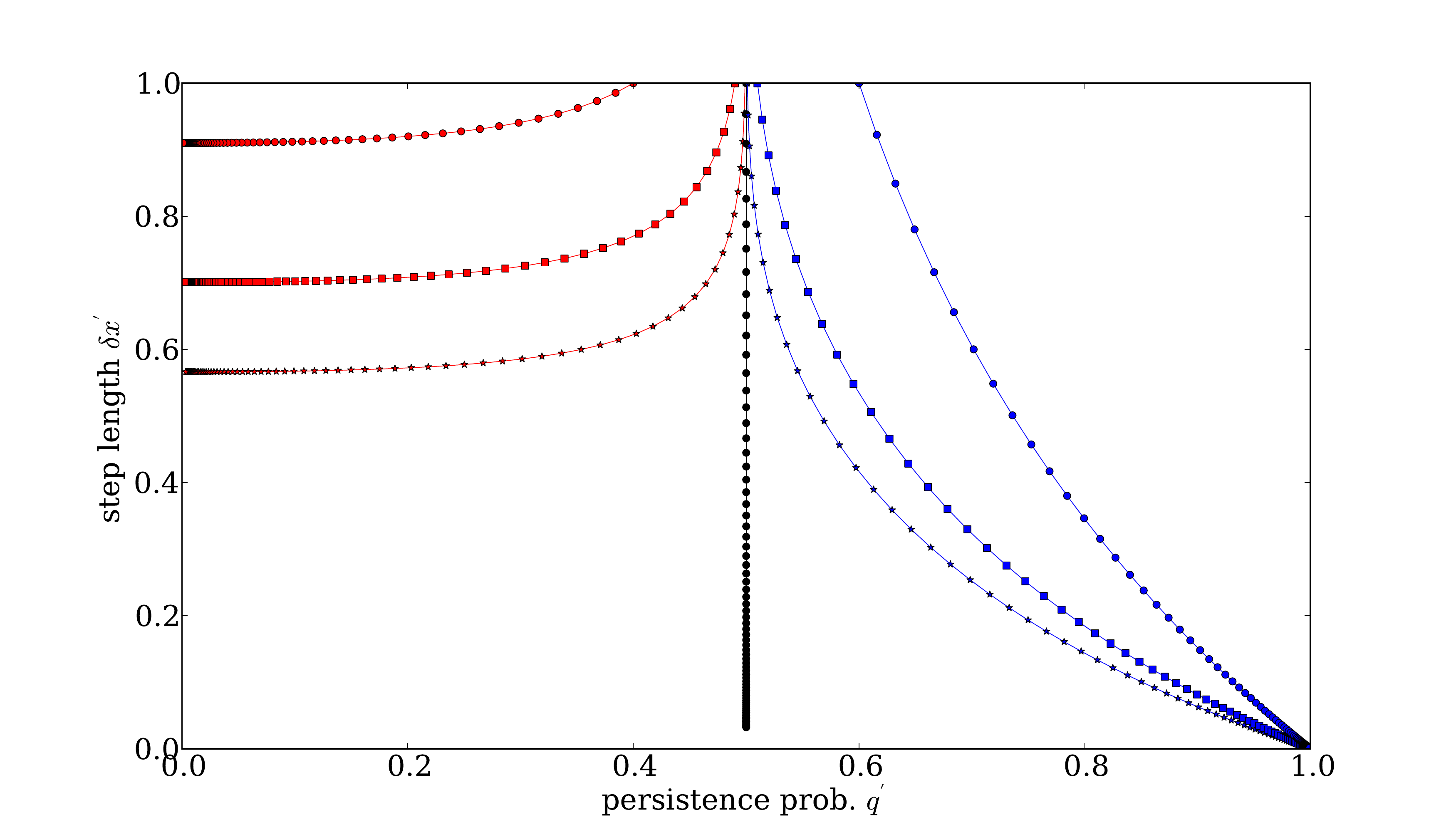}
\caption{\label{fig:flux2d} {\em 
Flux of parameter pairs $(q^{\prime},\delta x^{\prime})$ during the scaling transformation. The starting points, corresponding to scale parameter $s=1$, all have identical step length $\delta x=1$. With $s$ increasing from $1$ to $1000$, the step lengths $\delta x^{\prime}$ become smaller. For anti-persistent walks (red), the step lengths approach a constant asymptotic value, for persistent walks (blue), they decay to zero. The shown curves correpond to different initial $q$-values: red circles: $q=0.4$, red squares: $q=0.49$, red stars: $q=0.499$, black circles: $q=0.5$, blue stars: $q=0.501$, blue squares: $q=0.51$, blue circles: $q=0.6$.
}}
\end{figure}

We thereby have obtained two coupled scaling relations for the persistent probability and the step length: When the time step of a CDTRW is changed from $\delta t$ to $\delta t/s$, the new walk will have equivalent statistical properties if its parameter pair $(q,\delta x)$ is changed to $(q^{\prime}=q^{\prime}(q,s),\delta x^{\prime}=\delta x\cdot g(q,s))$, with the scaling functions $q^{\prime}(q,s)$ and $g(q,s)$ as defined above (For an example, see Fig.\ref{fig:flux2d}).

\begin{figure}[htb]
\includegraphics[width=15cm]{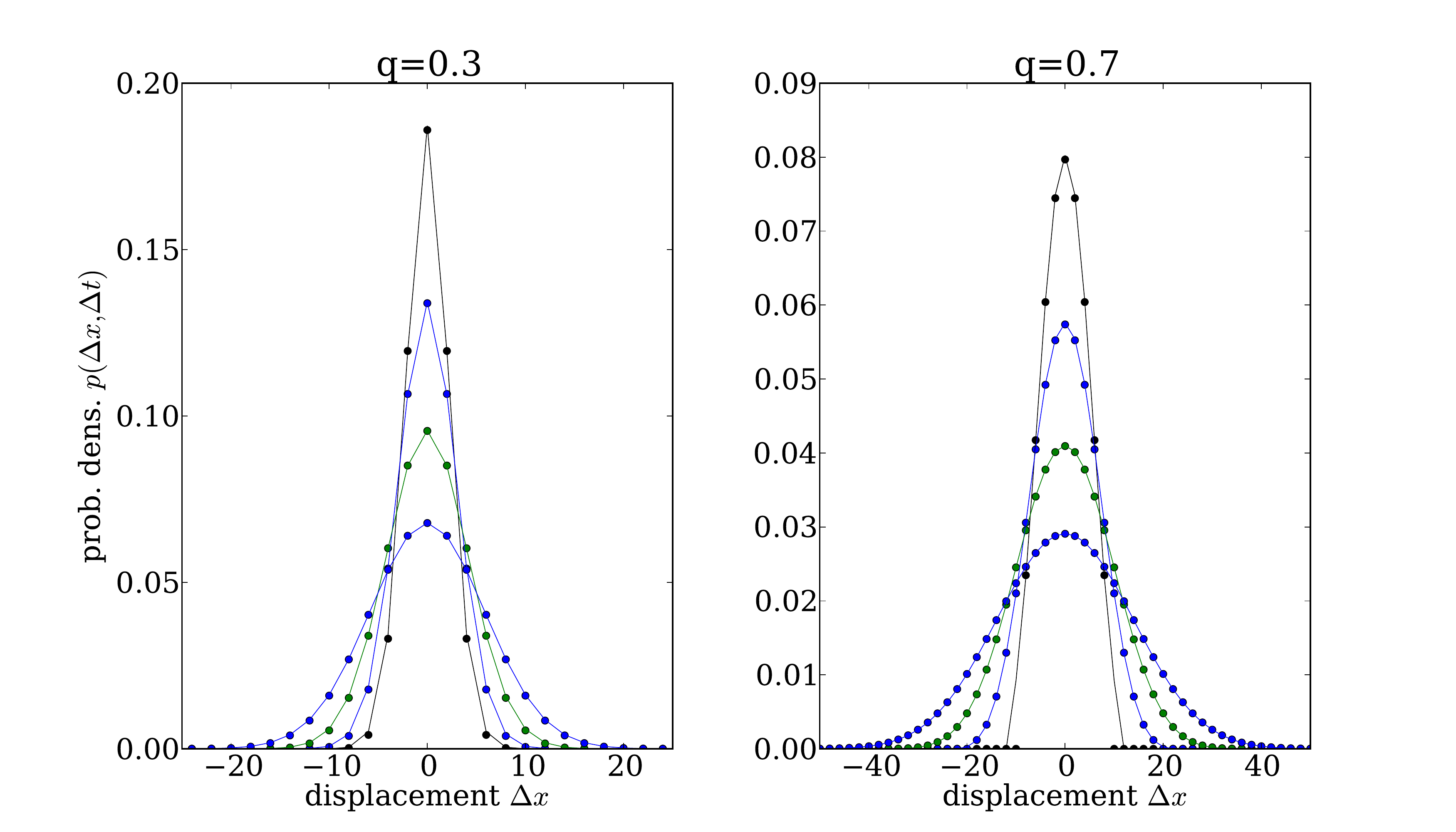}
\caption{\label{fig:swdAB} {\em 
Displacement distributions of CDTRWs for the persistence probabilities $q=0.3$ (left) and $q=0.7$ (right). Circles correspond to the `original' random walk, using a time step $\delta t=1$ and a step length $\delta x=1$. Lines correspond to the rescaled random walk, using a ten times smaller time step $\delta t=0.1$ and appropriately transformed values of the parameters $q$ and $\delta x$. The results are identical, thus confirming the scaling transformation.
}}
\end{figure}

We have tested the scaling relations by directly comparing the displacement distributions and the mean squared displacements for an original walk with sampling time $\delta t=1$ and a properly scaled walk with $\delta t=1/10$. The results are shown in Fig.\ref{fig:swdAB}.

\begin{acknowledgments}
This work was supported by grants from Deutsche Forschungsgemeinschaft.
\end{acknowledgments}


\bibliography{refs}


\end{document}